\documentclass[conference]{IEEEtran}
\usepackage{amsfonts,amsmath,latexsym,amssymb}
\usepackage{graphicx}
\usepackage{CJK}
\usepackage{multirow}
\usepackage{chemarrow}
\usepackage{setspace}
\usepackage{epstopdf}
\usepackage{flushend}
\begin{document}
\begin{spacing}{1}

\title{A Diffusion-Neuron Hybrid System for Molecular Communication}

\author{\IEEEauthorblockN{Peng He\IEEEauthorrefmark{1}, Yuming Mao\IEEEauthorrefmark{1}, Qiang Liu\IEEEauthorrefmark{1} and Kun Yang\IEEEauthorrefmark{1}\IEEEauthorrefmark{2}}
\IEEEauthorblockA{
\IEEEauthorrefmark{1}University of Electronic Science and Technology of China, Chengdu, China\\
\IEEEauthorrefmark{2}Network Convergence Laboratory, University of Essex, Colchester, UK\\
Email: hp6500@126.com, \{ymmao, liuqiang\}@uestc.edu.cn\\
kunyang@essex.ac.uk}\\
}

\maketitle

\begin{abstract}
Diffusion-based and neural communication are two interesting domains in molecular communication. Both of them have distinct advantages and are exploited separately in many works. However, in some cases, neural and diffusion-based ways have to work together for a communication. Therefore, in this paper, we propose a hybrid communication system, in which the diffusion-based and neural communication channels are contained. Multiple connection nano-devices (CND) are used to connect the two channels. We define the practice function of the CNDs and develop the mechanism of exchanging information from diffusion-based to neural channel, based on the biological characteristics of the both channels. In addition, we establish a brief mathematical model to present the complete communication process of the hybrid system. The information exchange process at the CNDs is shown in the simulation. The bit error rate (BER) indicator is used to verify the reliability of communication. The result reveals that based on the biological channels, optimizing some parameters of nano-devices could improve the reliability performance.
\end{abstract}


%
\IEEEpeerreviewmaketitle

\section{Introduction}
%
%
Molecular communication (MC), one of the most promising communication ways at nanoscale, develops rapidly due to great advance of the nanotechnology. The information is encoded in chemical molecules, transmitted via biological channels and exchanged between bio-inspired nano-devices [1]. In general, molecular communication could be divided into wireless and wired branches. Wireless MC is realized through diffusive propagation of the molecules, and could be further divided according to the communication distance and molecule species. Some typical examples includes bacterium, calcium signaling and pheromone. In wired MC, information is transmitted along some types of physical link, such as the neuron, blood vessel and microtubule.

Diffusion-based communication is an important branch of wireless MC, in which the molecules could only transfer slowly in a short distance. While nano-devices adopting diffusion-based communication could move freely in the environment rather than stay at a fixed position. Differently, neural communication, one typical example of the wired MC, is a fast and long range transmission way in human body. However, nano-devices of neural system lack flexibility, i.e., they have to stay at some key position to proceed the communication. For example, receiver should be close to the cell membrane, for monitoring the membrane potential during the communication. In nature, there are some cases that diffusion-based and neural ways cooperate to proceed the signal transmission, and we list two cases. For the first case, the Ca$^{2+}$ ions diffuse around the neurons, access the cell through Ca$^{2+}$ channels, impact the vesicle release activity to adjust the neural signaling. For the second case, some ions including Na${^+}$, K${^+}$, Cl${^-}$, diffuse and pass in and out the neuron through ion channels, altering the membrane potential to trigger the electric signal of neural fiber. In this paper, to share the different advantages and avoid shortages of the both communication ways referred before, we design an adjustable diffusion-neuron hybrid system by extending the biological background of the second case.

From the aspect of communication and mathematical scheme, abundant literatures investigate the two communication ways discretely, which lay a theoretical basis for our research. Such as [2], [3], [4], respectively discuss the channel capacity, transmission reliability and delay for controllable diffusion-based MC. About the neural communication, in [5], a mathematical model of single neuron at Cornu Ammonis region is established. [6] analyzes the error probability and delay of neural transmission. In [7], transmission interference in neural channel is studied.

However, with the aid of those basis, it's still difficult to establish a hybrid communication architecture. In traditional communication, a gateway device is necessary for connecting the wired and wireless networks. An typical example is the home router, with wired and wireless interfaces integrated in one device, coping with the information exchange between different mediums. At nanoscale, a similar device is needed to realize the same function in the hybrid architecture, which is named as connecting nano-device (CND) in this paper. The major challenge is a reasonable design of the CND, and a feasible mechanism to exchange signal from diffusion-based channel to neural channel.

\begin{figure*}[!t]
\centering
\includegraphics[width=5in]{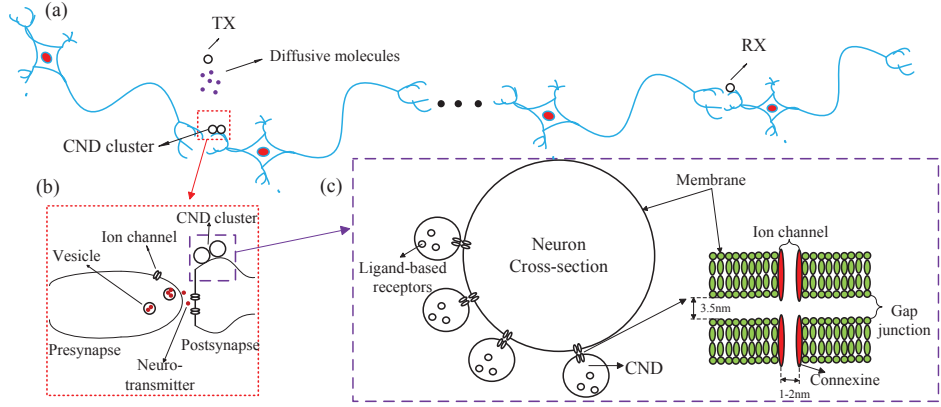}
\caption{Biological model of the hybrid communication system.}
\end{figure*}

In this paper, we propose a diffusion-neuron hybrid communication architecture, in which the information is transmitted from diffusion-based to neural channel, not including the reverse process. Different from [5]-[7], in our work, we consider the general controllable diffusion process between nano-devices, not limited by the ion diffusion of the natural neural activity. Due to the restriction of size and power for nanoscale device, a CND cluster is utilized rather than a single CND to perform the task. We define two interfaces for CNDs, named diffusion interface and neural interface, respectively coping with the communication for different mediums. We design the practice function of the CNDs, which apply the electric current stimuli to trigger the neural signaling. This method is utilized in [8], [18], [19] and [20], for design of the nano-device in neural communication. [8] proposes the equivalent neural devices method, that the multiple neural devices are uniformly driven by one current. While in our paper, for the neural interface of each CND, we apply the different strength of currents, that is inspired according to signaling of the diffusion interface. We describe it as the current inspired mechanism, through which the signal is exchanged from diffusion-based to neural channel. We discuss this mechanism in biological method and present it in the complete communication process with mathematical model. The expression of bit error rate (BER) is given, to study the reliability of the hybrid model. In the simulation, we study how the information is exchanged at the CNDs, as well as which parameters impact on the BER performance. Our work provides some guidance for designing the hybrid communication architectures at nanoscale, contributing to the development of body area network.

The rest of the paper is organized as follows. In Section II, we introduce the biological background of diffusion and neural communication. Based on this, the entire hybrid system is described, the CND scheme is discussed. The mathematical model of a complete communication process is established in Section III. After that, the simulation result is presented in section IV. Finally, we conclude the paper in section V, as well as the future extension of this work.

\section{Biological Method}
Natural diffusion phenomena exist abundantly around the neural system. Such as the ions Ca$^{2+}$, Na${^+}$, K${^+}$, Cl${^-}$, diffuse around the neurons, and play an important role in normal activity. Though they are parts of the neural signaling, the diffusion behaviour could be extended into general diffusion-based communication by applying more types of molecules in artificial nano-devices, to implement the interface of neural network to other MC networks. The neural transmission is rather complex, proceeded among billions of neurons. The typical communication process in a single neuron includes neural firing, action potential (AP) transmission, vesicle release and postsynaptic response. Briefly, information is encoded in AP spike trains, which are generated in process of neural firing. The potential rises and falls rapidly like water wave, moves along the neural fiber until arriving presynapse, as Fig. 1 (b) shown. Then, ion channels are open, letting Ca$^{2+}$ ions come in and promoting the process of vesicle release. Neurotransmitters stored in vesicles are released, and propagate in the synaptic gap before arriving in postsynapse, where the postsynaptic response happens to raise the postsynaptic membrane potential, due to the exchange of various ions. Postsynaptic response potential is divided into excitatory postsynaptic Potential (EPSP) and inhibitory postsynaptic Potential (IPSP), both of which are much lower than one AP. The former contributes the neural firing process while the latter restrains. With the effect of EPSPs and IPSPs, when membrane total potential exceeds a threshold in postsynaptic response, an AP spike is generated, followed by a period of resting time, named absolute refractory period in neuroscience. During the absolute refractory period, no AP spikes could be generated.

The hybrid system designed is composed of one transmitter (TX), one receiver (RX) and multiple connecting nano-devices (CND), across both the diffusion-based and neural communication mediums. The biological model is shown in Fig. 1 (a). TX is a biological nano-device as most literatures desire, which could produce, emit and receive specific type of molecules. RX considered in this paper is the typical nano-device in neuron communication, with ability of monitoring membrane potential and decoding information. CND cluster are designed as the relaying bridge, which could receive molecules from diffusion-based channel and evoke EPSPs to fire the AP spike trains in the neural channel.

As a key element of the hybrid system, CND cluster should be able to cope with the communication both in diffusion-based and neural channel. The CND cluster around a neuron is shown in Fig. 1 (c). They are bio-inspired nano-devices containing diffusion and neural interface. Responsible for communicating with TX, the diffusion interface is formed by plenty of biological receptors, which are embedded on the membrane surface, distributed uniformly to receive signal molecules in a small receiving space $V_r$. We utilize the ligand receptors for diffusion-based interference described in [9]. Molecules are bound and released with different rates, when they are in the receiving space. We adopt the design of [8] for the neural interface. CNDs should be able to produce tiny biological electricity to stimulate the neural firing process. An example of generating biological electricity is introduced in [17], depending on oxidation of organic substrates, which is promising to implement in an artificial cell. The neural membrane and CND are directly contacted with gap junctions, which is also the connection way of two adjacent neurons. Clusters of connexine proteins (named Cx36), combine to form an ion channel, allowing the ions and tiny electricity passing through during the neural encoding process [10]. The gap junction method is supported by neuroscience in [11] and [12], where it is studied in oscillatory behaviors and synchronization phenomena between neurons.

The on-off key (OOK) modulation of diffusion-based MC is used at TX, determining that CNDs could receive much molecules for bit "1" and few molecules for bit "0". The idea of current inspired mechanism is that according to quantity of the received molecules per bit, CNDs operate independently and inspire the electric currents of various strengths, to implement the device-based neural firing process. First of all, to avoid the possible interference, this process should proceed when there is no normal neural activity. Result of [18] indicates that different shape of current waves make difference for neural firing and the sine wave is the best. The current of several $\mu A/cm^{-2}$ is enough to fire an AP. Hence, in this paper, we consider the sine wave current with various amplitudes. During this process, there is no decoding or complex calculation for CNDs. The sine wave amplitude just varies with the received molecules quantity per bit in different time slots. Then, different strength of currents evoke the EPSPs with various amplitudes, which contribute to raise the membrane total potential. As described before, the AP spikes could be generated only when membrane total potential is strong enough to exceed the threshold. Hence, the binary sequence information is encoded in the AP spike trains (no spikes = "0", spikes = "1"), transmitted in the neural network until decoded by RX.

\begin{figure}[!t]
\centering
\includegraphics[width=3.5in]{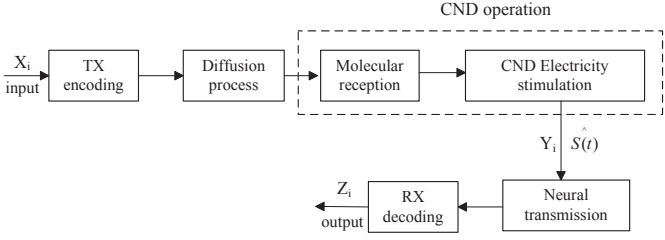}
\caption{Communication process diagram of the hybrid system.}
\end{figure}

\section{Mathematical model}
To detail the design of the hybrid system described above, we present the communication process from TX to RX with a brief mathematical model. As Fig. 2 shown, the communication process is divided into 5 parts, which are transmitter encoding, diffusion process, CND operation, neural transmission and receiver decoding respectively. We describe it as follows: an input binary information sequence, denoted by $X_i,i=1,2,...$, is encoded in molecules by TX and emitted into the medium. Those molecules diffuse freely until received by CNDs. Then, CNDs stimulate the membrane of the neuron cell through tiny electronic currents, firing the AP spike trains $\hat{S(t)}$, in which relaying binary sequence $Y_i$ is coded. After a distance of neural transmission, the AP trains are decoded to $Z_i$ by RX through monitoring the membrane potential.
\subsection{Transmitter Encoding}
In this paper, we only consider one transmitter. One time slot is utilized to transmit one bit, with fixed length of $T$. Let $R(0,t)$ be the molecular emitting rate of transmitter, at location of 0 and time of $t$. The OOK modulation is adopted to encode the input signal $X_i$, meaning that at the beginning of each time slot, quantity of $Q_{tx}$ molecules are emitted to indicate bit "1", i.e.,
\begin{equation}
R(0,t)=
\begin{cases}
Q_{tx}& \text{bit "1"}\\
0& \text{bit "0"}
\end{cases}
\end{equation}

\subsection{Diffusion Process}

Molecules diffuse in the environment and perform the brownian movement. For the diffusion process of single source, the molecular concentration follows the Fick's second law, with the solution of the Green function [1],

\begin{equation}
c(x,t) = R(0,t)\cdot g(x,t)+n_1(t)
\end{equation}
Where $c(x,t)$ is the molecular concentration at location $x$ and time $t$. $n_1(t)$ is the gaussian noise of diffusion channel, s.t., $n_1(t)\sim N(0,\sigma_1^2)$. $g(x,t)$ is the Green function mentioned above, expressed as $g(x,t)=\frac{1}{(4\pi Dt)^{3/2}}exp(-\frac{x^2}{4Dt})$. Here $D$ is the diffusion coefficient related with molecule species and environment condition. When $t=kT$, there have been $k$ bits sent from the transmitter. Let $r$ be the distance between TX and a CND, the concentration at a CND is accumulated with time and expressed as,

\begin{equation}
c(r,t) = \sum_{i=1}^k R(0,t-iT)\cdot g(r,t-iT)+n_1(t)
\end{equation}

\subsection{CDN Operation}
Assuming there are $M$ CNDs in the cluster. Let $r_k$ be the distance between transmitter and $k^{th}$ CND. Utilizing the ligand-based receptors for diffusion interface, according to (3), for $i^{th}$ bit at $k^{th}$ CND, we calculate the quantity of received molecules as,

\begin{equation}
Q_{rx}(k,i)=\frac{\epsilon_1\cdot V_r\cdot\rho}{\epsilon_{-1}}\cdot \int_{iT}^{(i+1)T} c(r_k,t)dt
\end{equation}
Where $\epsilon_1$ and $\epsilon_{-1}$ respectively means the binding and release rate of the molecules [9]. $\rho$ is the concentration of the ligand-based receptors on surface of CNDs. $V_r$ is the receiving space of one CND. Based on the current inspired mechanism in section II, electric current $A_k(t)$, inspired by $k^{th}$ CND, are produced according to knowledge of $Q_{rx}(k,i)$. So, we have,

\begin{equation}
A_k(t)=f(Q_{rx}(k,i)), \text{$iT\leq t<(i+1)T$}
\end{equation}
Where $f(x)$ is a positive relation function of $x$, indicating the current inspired mechanism. During the $i^{th}$ time slot, more molecules received, stronger the electricity should be inspired. Applying sine wave form, we rewrite (5) as,

\begin{equation}
A_k(t)=\mu\cdot Q_{rx}(k,i)\cdot sin(at+b)
\end{equation}
Where $\mu$ indicates the sensitivity of the current inspired mechanism, meaning the current strength per molecule. $a$ is the fundamental frequency of the sine wave. $b$ means the phase, which is usually set as 0.

With the given electric currents from all CNDs, the postsynaptic potential $V(t)$ could be obtained according to Spike Response Model, described in [13],

\begin{equation}
V(t)=V_0 + \sum_k^M \int_{t_k}^{\infty} \nu(s) A_k(t-s)ds
\end{equation}
Where $V_0$ is the resting potential without any neural activity. The current of $k^{th}$ CND is generated at time of $t_k$. $\nu$ is the linear response of the cellular membrane to an input pulse of current, given that the neuron did not fire in the recent time. $A_k(t)$ means the current fired by $k^{th}$ CND. The AP spike trains are generated only when $V(t)$ is strong enough to exceed the firing threshold $\theta_1$, the typical form of the AP spike trains is given by,

\begin{equation}
\hat{S(t)}=\sum_j \psi(t-\hat{t_j})
\end{equation}
Where $\psi(t-t_j)$ means the shape of one spike, which is triggered at time $\hat{t_j}$ if $V(t)=\theta_1$ and $\frac{dV(t)}{dt}>0$. It reaches the maximum after a short time and quickly drop, followed by absolute refractory period of over 15 ms, preparing generating the next AP spike. During $i^{th}$ time slot, if at least one AP spike $\psi$ exists in $\hat{S(t)}$, $Y_i$ is "1", else $Y_i$ is "0".

\subsection{Neural Transmission}

\begin{figure}[!t]
\centering
\includegraphics[width=3.5in]{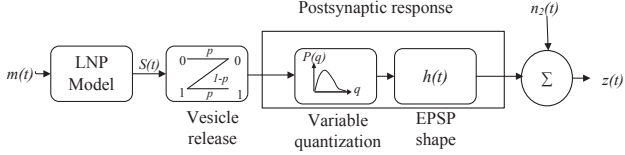}
\caption{Transmission diagram of a single neuron}
\end{figure}

Inspired by CNDs, signal should pass plenty of neurons before decoded by RX. The transmission could be regarded as the repeated process of one single neuron transmission. A mathematical model of a single neuron communication system is established in [5], shown in Fig. 3. $m(t)$ is the stimuli signal for neural firing, regarded as the input. LNP model contains the process of neural firing and AP transmission, that is introduced in section II and illustrated detailedly in [14]. $S(t)$ is the AP spike trains transmitted in this neuron. Vesicle release is a probabilistic model. Neurotransmitters release from the vesicles, with probability of $p$, which is related with the neuron types. In postsynaptic response block, the random value $q$, with the probability density $P(q)$, indicates the variability of the EPSP amplitudes. EPSP shape is the second step of the postsynaptic response, moulding the voltage by alpha function $h(t)$, expressed as,

\begin{equation}
h(t)=\frac{h_p}{t_p}\cdot t\cdot exp(1-\frac{t}{t_p})
\end{equation}
Where $h_p$ is the EPSP magnitude and $t_p$ indicates corresponding time of magnitude peak. $n_2(t)$ is the neural noise in the postsynaptic potential, modeled as additive gaussian noise, s.t., $n_2(t)\sim N(0,\sigma_2^2)$. The output signal $z(t)$, which is also the stimuli signal of the next neuron, is formed at the postsynapse.

\subsection{Receiver Decoding}
The decoding process is monitoring the membrane potential to obtain the AP spike trains $\hat{S(t)}$. On general condition, neural transmission is an efficient way with extremely low error probability because neurons has ability to eliminate interference [15]. So, $\hat{S(t)}$ could be seen undistorted when decoded by RX. Described in [16], it is efficient to compare a threshold $\theta_2$ to correlation $\gamma$, between membrane potential and $h(t)$. The larger probability to inspire the APs, the bigger $\gamma$ is. For $i^{th}$ time slot, $iT \leq t<(i+1)T$, output $Z_i$ could be determined as,

\begin{equation}
Z_i=
\begin{cases}
1& \text{$\gamma\geq \theta_2$}\\
0& \text{$\gamma<\theta_2$}
\end{cases}
\end{equation}

The reliability of the system is verified by indicator of the bit error rate (BER). Based on the communication process diagram in Fig. 2, we give the BER expression of the hybrid system as,

\begin{equation}
P_e=pr(Z_i=0|X_i=1)+pr(Z_i=1|X_i=0)
\end{equation}

Considering the middle binary sequence $Y_i$, we have,

\begin{equation}
\begin{split}
P_e=& pr(Y_i=0|X_i=1)pr(Z_i=0|Y_i=0)\\
&+pr(Y_i=1|X_i=1)pr(Z_i=0|Y_i=1)\\
&+pr(Y_i=0|X_i=0)pr(Z_i=1|Y_i=0)\\
&+pr(Y_i=1|X_i=0)pr(Z_i=0|Y_i=1)
\end{split}
\end{equation}

Note that $pr(Y_i|X_i)$ is related with TX encoding, diffusion process and CND operation. Differently, $pr(Z_i|Y_i)$ is related with neural transmission and receiver decoding. We could see that for the diffusion and neuron sides, if information of the two sides are transmitted both accurately or both wrong, then the final decoding of receiver is right. Else, the receiver will get the error bit.

\begin{figure*}[!t]
\centering
\includegraphics[width=7in]{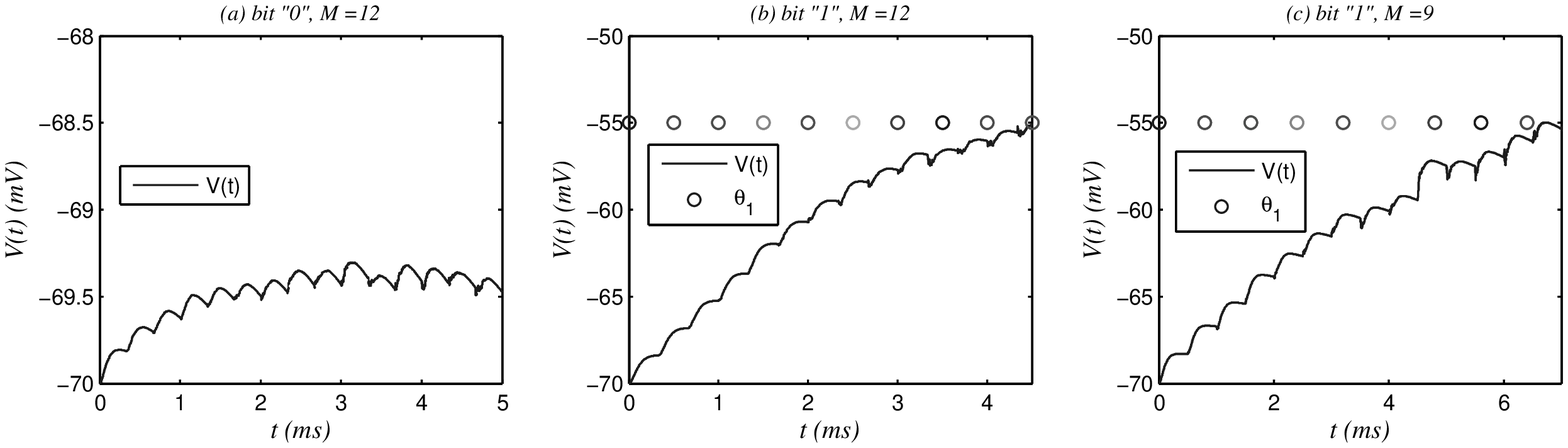}
\caption{Neural firing for various bits, inspired by multiple CNDs,}
\end{figure*}

\section{Numerical Result}
In this section, we simulate diffusion-neuron hybrid system detailed in section II and III. We particularly focus on the signal transform process from diffusion-based to Neural channel. Some controllable nano-device setting parameters, including CND quantity $M$, CND sensitivity $\mu$, CND fundamental frequency $a$, time slot $T$ and receiver threshold $\theta_2$, are investigated in terms of their impact to BER performance. The goal of the simulation is to test the signal exchange process in CND operation, and show that optimizing those parameters could improve the communication reliability.

\subsection{Parameter Setting}

The results are computed under a common set of parameters, which are set as follows. The diffusion-based coefficient $D$ is set as 0.1$\mu m^2/s$. Molecule quantity of bit "1" $Q_{tx}$ is fixed as $10^5$ [3]. Length of one time slot $T$ is set from 100 to 300ms. Distances $r_k$ between TX and different CNDs is set in range of 15-20 $\mu m$. $\epsilon_1$ and $\epsilon_{-1}$ are set as 0.1 and 0.08. $\rho$ is set as 0.5 $\mu$mol/liter, similar with [9]. The quantity of CNDs $M$ changes from 5 to 12. electricity per receiving molecule $\mu$ is set from 20 to 40 $nA/cm^{-2}$ per molecule. The fundamental frequency $a$ changes from 40 to 80 Hz. $b$ is fixed as 0. The initial resting membrane potential $V_0$ is set as -70mV for a typical neuron, with corresponding firing threshold $\theta_1$ as -55mV. Receiver decoding threshold $\theta_2$ is set as 1 similar with [4]. EPSP magnitude $h_p$ and time to peak $t_p$ is respectively 1mV and 0.5ms when calculating correlation $\gamma$. The variance of diffusive noise $\sigma_1$ and neural noise $\sigma_2$ are set as $0.1/\mu m^3$ (average one molecule in 10 $\mu m^3$) and $0.1 mV$.

\subsection{CND Operation}
We simulate the CND operations in computer according to the well known Hodgkin Huxley model [13], and present the signal exchange process from the diffusion-based to neural channel, which is the key of the hybrid system.

In Fig. 4, We show the neural firing process at the CDNs. Each small wave curve means one EPSP contribution from CNDs. They have various amplitudes, decided by quantity of the received molecules. We can see that the potentials vary from -70mV, for bit "1" and "0", there is an obvious difference. Membrane Voltage of bit "0" increase slowly in Fig. 4 (a), the increase speed even falls behind the EPSP decreasing. So the potential could not reach $\theta_1$ to fire an AP. In Fig. 4 (b) and (c), the potential could reach $\theta_1$ because CNDs provide currents of enough strength. But potential of Fig. 4 (b) increase much faster, reach $\theta_1$ by only about 4.4ms. In Fig. 4 (c), it needs nearly 6.7ms to fire. The reason is that for a larger $M$, more EPSPs will be generated to promote the neural firing and save time.

\begin{figure}[!t]
\centering
\includegraphics[width=3in]{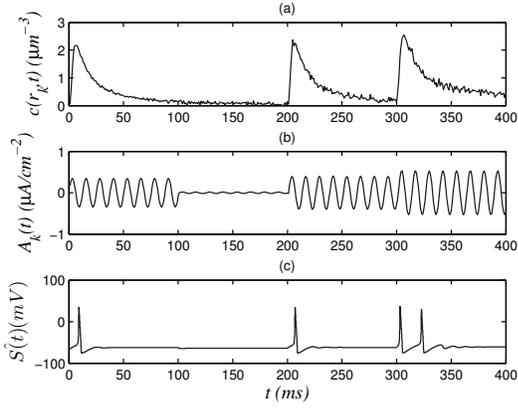}
\caption{(a) molecular concentration $c(r_k,t)$, (b) electric current $A_k(t)$, (c) AP spike trains $\hat{S(t)}$, for bit "1011", $T$ = 100ms, $r_k$ = 15$\mu m$, $\theta_2$ = 1,$M$ = 6, $\mu$=20$nA/cm^{-2}$ per molecule.}
\end{figure}

Fig. 5 presents the signal exchange process of the bit "1011" at the CNDs. The molecular concentration $c(r_k,t)$, the corresponding inspired current $A_k(t)$ and the AP spike trains $\hat{S(t)}$ are respectively shown. 400ms is divided into 4 time slots, in which 4 bits are transmitted. In Fig. 5(a), the curve is full of sawteeth which are caused by noise. We could see that, during the time from 100 to 200ms, low $c(r_k,t)$ means receiving little molecules for this bit, and the current inspired from CNDs is not enough to fire an AP spike. Thus this bit is "0". From 300 to 400 ms, there are two AP spikes, because the molecules of front bit and the noise make CND receive more molecules than other bits, inspiring a stronger current with the amplitude of about 0.6 $\mu A/cm^{-2}$, to fire more than one spike, which also means one bit "1".

\begin{figure}[!t]
\centering
\includegraphics[width=3in]{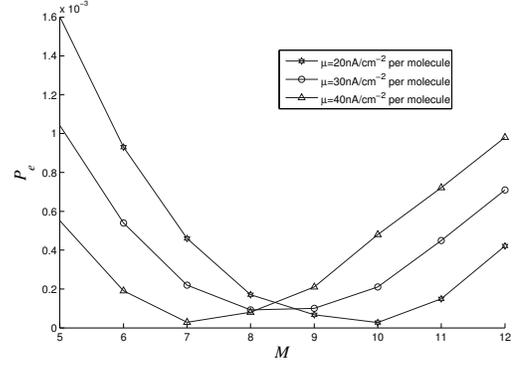}
\caption{$f(Q_{rx}) = \mu\cdot Q_{rx}\cdot sin(at+b)$, $P_e$ for different M and $\mu$, $T$ = 100ms, $\theta_2$ = 1.}
\end{figure}

\begin{figure}[!t]
\centering
\includegraphics[width=3in]{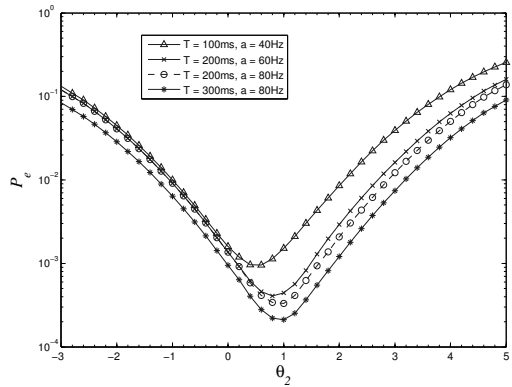}
\caption{$P_e$ varies with $T$ and $a$, under different $\theta_2$, $M$=6, $\mu$=20$nA/cm^{-2}$ per molecule.}
\end{figure}

\subsection{System Reliability}
Fig. 6 indicates the BER performance under various $M$ and $\mu$. Increasing $M$ will propel the BER to fall first and rise then. The reason is two-folds. On one hand, more CNDs applied means more received molecules for bit "1", correspondingly larger probability to fire the AP spikes in $\hat{S(t)}$. On the other hand, caused by channel noise and residual molecules of front bits, increasing CNDs quantity $M$ will also increase the probability to fire the AP spikes for bit "0", and the spikes of bit "1" may also extend into bit "0" to lead the error decoding. Alteration of parameter $\mu$ will change the sensitivity of the CNDs. We could see that under larger $\mu$, fewer CNDs are needed to achieve the lowest BER. The result suggests that for different quantity of CNDs, $\mu$ should be adjusted to achieve a lower BER.

In Fig. 7, the relationship between BER, $T$ and $a$ under different $\theta_2$ is revealed. We could see that, under a fixed setting of TX and CNDs, RX parameter $\theta_2$ impacts BER directly. When $\theta_2$ stays around 1, the decoding process could achieve minimum BER. For a longer $T$, the interference of neighbouring bit is smaller in whether diffusion-based or neural channel, which leads decrease of the BER. But the corresponding cost is a longer transmission delay. Also in range of 40 to 80Hz, increase the CND frequency may lead more AP spikes be produced and decrease the error probability, but higher CND working frequency requires more for nano-device manufacture.

\section{Conclusion}
In this paper, we develop a hybrid system containing diffusion-based and neural communication channels. We define the function of the connected nano-devices, which are used to connect the both channels. The detail mechanism of exchanging information from diffusion-based to neural channel are proposed, and the complete communication process is studied in the mathematical model. The results show the detail process of the mechanism, indicating that it's feasible to realize the communication in the hybrid system. Optimizing some parameters of the nano-devices could achieve lower BER performance for this hybrid system. Our research could be extended to the design of communication across various biological mediums in body area network.

Although this work mainly focuses on the system model and information exchange mechanism of single direction, the future work includes the reverse direction, i.e., communication from neural to diffusion-based channel, as well as the analysis of the channel capacity, transmission delay for the hybrid system.

\setcounter{equation}{0}
\newcommand\threequation{}
\appendices


\section*{Acknowledgment}

This work in the paper was supported by National Natural Science Foundation of China (NSFC-2014-61471102) and EU FP7 Project CLIMBER (GA-2012-318939).



%

\end{spacing}
\end{document}